\documentclass[lettersize,journal]{IEEEtran}
\usepackage{amsmath,amsfonts}

\usepackage{graphicx}
\usepackage{bm}
\usepackage[dvipsnames]{xcolor}
\usepackage{ifthen} 
\usepackage[shortlabels]{enumitem}
\usepackage{cite}
\usepackage{mathtools} 
\usepackage{multirow}
\usepackage{hhline}
\usepackage{caption}
\usepackage{stfloats}
\usepackage{subcaption}
\usepackage{array}
\usepackage{booktabs}
\usepackage{setspace}

\def\mylist#1 {\ifx!#1\else\makebox[4em][r]{#1} \expandafter\mylist\fi}

\newcommand{\mb}[1]{\mathbf{#1}}
\newcommand{\tu}[1]{\textup{#1}}

\newcommand{\Herm}[0]{\textup{H}}
\newcommand{\Hmsp}[0]{{\mspace{1mu}\textup{H}}}
\newcommand{\F}[0]{\bm{\mathcal{F}}}

\newcommand{\dInt}[1]{\text{d}\mspace{-2mu}{#1}}
\newcommand{\figwid}[0]{0.98}
\newcommand{\R}[0]{\mathcal{R}}

\newcommand{\Fnc}[0]{\mathbb{F}}

\newcommand{\Ii}[0]{\bm{\mathcal{I}}}
\newcommand{\sinc}[0]{\textup{sinc}}
\newcommand{\ypre}[0]{{\mathbf{y}}'}
\newcommand{\ypref}[0]{{\mathbf{y}}'_\textup{f}}
\newcommand{\Xpref}[0]{{\bm \Phi}'_{\textup{f}}}

\newcommand{\XprefH}[0]{{\bm \Phi}{'}^\textup{H}_{\mspace{-6mu}\textup{f}}}

\newcommand{\subf}[1]{{#1_{\textup{f}}}}
\newcommand{\subp}[1]{{#1_{\textup{p}}}}
\newcommand{\rhoc}[0]{\hat{\bm{\rho}}}
\newcommand{\sigvdvd}[0]{\bm{\Sigma}_{\bar{\mb v}'_{\mspace{-1.5mu}\textup{f}}\bar{\mb v}'_{\mspace{-1.5mu}\textup{f}}}}

\newcommand{\intvl}[0]{\lambda_{\textup p}}

\DeclareMathOperator*{\argmin}{arg\,min}

\newcommand{\Seta}[0]{\hspace{2pt}\hat{\mskip-3.6mu\bm{\mathcal{S}}}_{\mspace{-2mu}\eta}}	
\newcolumntype{P}[1]{>{\centering\arraybackslash}p{#1}}

\renewcommand{\arraystretch}{1.1}

\begin{document}
\title{Multicarrier Spread Spectrum Communications with Noncontiguous Subcarrier Bands for \\HF Skywave Links}
\author{\IEEEauthorblockN{Brandon T. Hunt\IEEEauthorrefmark{1}, Hussein Moradi\IEEEauthorrefmark{2}, Behrouz Farhang-Boroujeny\IEEEauthorrefmark{3}}\\
    \IEEEauthorblockA{\IEEEauthorrefmark{1}Electrical Engineering Department, \textit{Montana Technological University}, bhunt@mtech.edu} \\
    \IEEEauthorblockA{\IEEEauthorrefmark{2}Wireless Research Team, \textit{Idaho National Laboratory}, hussein.moradi@inl.gov} \\
    \IEEEauthorblockA{\IEEEauthorrefmark{3}ECE Department, \textit{The University of Utah}, farhang@ece.utah.edu}
\thanks{This manuscript has been authored by Battelle Energy Alliance, LLC under Contract No. DE-AC07-05ID14517 with the U.S. Department of Energy. The United States Government retains and the publisher, by accepting the article for publication, acknowledges that the United States Government retains a nonexclusive, paidup, irrevocable, world-wide license to publish or reproduce the published form of this manuscript, or allow others to do so, for United States Government purposes. STI Number: INL/JOU-23-75226}}

\maketitle

\begin{abstract}
Growing traffic over the high-frequency (HF) band poses significant challenges to establishing robust communication links. While existing spread-spectrum HF transceivers are, to some degree, robust against harsh HF channel conditions, their performance significantly degrades in the presence of strong co-channel interference. To improve performance in congested channel conditions, we propose a filter-bank based multicarrier spread-spectrum waveform with noncontiguous subcarrier bands. The use of noncontiguous subcarriers allows the system to at once leverage the robustness of a wideband system while retaining the frequency agility of a narrowband system. In this study, we explore differences between contiguous and noncontiguous systems by considering their respective peak-to-average power ratios (PAPRs) and matched-filter responses. Additionally, we develop a modified filter-bank receiver structure to facilitate both efficient signal processing and noncontiguous channel estimation. We conclude by presenting simulated and over-the-air results of the noncontiguous waveform, demonstrating both its robustness in harsh HF channels and its enhanced performance in congested spectral conditions. 

\end{abstract}


\section{Introduction}\label{sec:introduction}
\IEEEPARstart{D}{ue} to the uncoordinated use of the high-frequency (HF) band, transmissions are commonly polluted with narrowband interferers. To maintain communications in the presence of a narrowband interferer, multicarrier spread spectrum systems can be used \cite{HaraPrasad1997,Kondo1996,FarhangFurse2004,wasden,sibbett_nmf}. Such systems take advantage of increased spectral diversity to remain operable at low signal-to-interference-plus-noise ratios (SINRs). The improved performance of these spread-spectrum systems stems directly from the increased bandwidths they occupy \cite{wasden}. 

However, increasing signal bandwidths also increases the chance of exposing signals to additional interferers. Several studies (e.g., \cite{hf1363, warner_cad, furmanHf13, bergHfia}) report that sufficient bandwidth is frequently unavailable to transmit these wideband waveforms without coinciding with other users. Further, when multiple interferers are present across the transmission band, wideband systems are outperformed by narrowband systems which are able to avoid the most prominent users \cite{andersson95}. 

Hence, whereas wideband systems are robust against a limited amount of co-channel interference, avoiding interference altogether provides the best performance when operating in congested environments. To combine the advantages of both narrowband and wideband systems, we propose an extension to the family of multicarrier spread-spectrum waveforms in which the signal energy is allocated over a set of noncontiguous frequency bands. By separating the frequency bands, the signal propagates through a greater diversity of channel conditions while gaining the ability to adapt to spectral activity in the transmission band. 

To underscore the motivation of developing a noncontiguous waveform, we note the following potential benefits:
\begin{enumerate}[(1),label=\roman*)]
\item
\textit{Channel diversity:} Spreading the subcarriers over a broader bandwidth increases spectral diversity and may lead to a more stable communication link. While a deep channel fade may affect several adjacent subcarriers in a conventional multicarrier spread-spectrum system, such a phenomenon is less likely when the signal is spread over a wider bandwidth \cite{laraway_milcom16}.
\item
\textit{Time and frequency stability:} In many channels, the channel quality may vary significantly and unpredictably across time and frequency; e.g., in skywave HF links, channel conditions are dependent on factors such as the time of day and solar activity \cite{goodman_space, johnson2012}. Hence, by spreading the transmitted signal over a broad spectral range, there is an increased chance of recovering a greater number of the subcarrier bands, particularly when broadcasting to multiple scattered nodes.
\item
\textit{Dynamic spectral usage:} When transmitting wideband signals, there may be no sufficiently unoccupied spectrum to establish a communications link \cite{hf1363, warner_cad, furmanHf13}. A noncontiguous system is able to place its subcarriers between other users by bonding nonadjacent subchannels, thereby exploiting narrowband portions of unused spectrum.
\item
\textit{Multiple access:} The unrestricted placement of subcarriers provides a means for multiple access. This may be done by assigning different subcarriers to different users.
\end{enumerate}

\begin{figure}[t]
	\centering
	\small{C-FMT-SS}
	
	{\includegraphics[width = 0.8\columnwidth]{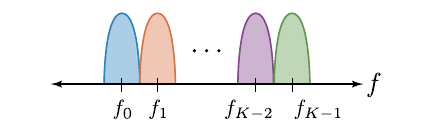}\label{subfig:cfmtss}}\\ \vspace{8pt}
	
	\small{NC-FMT-SS}
	
	{\includegraphics[width = 0.8\columnwidth]{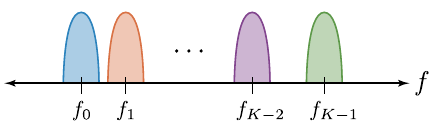}\label{subfig:ncfmtss}}
	\caption{Example spectra of contiguous and noncontiguous FMT-SS with $K$ subcarriers.}
	\label{fig:FMTspectrum}
\end{figure}

The noncontiguous transceiver system we develop in this work is based on the filter-bank multicarrier spread-spectrum (FBMC-SS) waveform. FBMC-SS is chosen for its improved rejection of partial-band interference as compared to other modulations such as OFDM-based spread spectrum \cite{HaraPrasad1997, Kondo1996, FarhangFurse2004, wasden}. We use a filtered-multitone (FMT) filter-bank waveform which yields a simple receiver structure and provides trivial interference suppression \cite{FMT1, FMT2, sibbett_nmf}. We refer to the contiguous and noncontiguous variants of the FMT spread-spectrum waveforms as C-FMT-SS and NC-FMT-SS, respectively; see Fig.~\ref{fig:FMTspectrum}. The C-FMT-SS waveform is equivalently FBMC-SS which has been studied in \cite{Kondo1996, wasden, haab_nmf}, among others, with the abbreviation C-FMT-SS being used herein to contrast against its noncontiguous variant.

Note that if the developed noncontiguous system was not spread-spectrum, the received signal could be treated as a collection of several narrowband transmissions and simply processed accordingly. However, the system proposed herein is \textit{spread spectrum}, and accordingly, each subcarrier conveys a piece of the same data symbol in the form of a chip. The challenge to the transceiver system thus lies in recombining the chips from the separated subcarriers in a way that maximizes the processing gain at the receiver output.

Finally, to develop a noncontiguous system, several changes must be made to both the transmitter and the receiver to accommodate the wide signal bandwidth of the resulting waveform. These changes must be considered in both the software elements (e.g., developing efficient processing structures) and the hardware (e.g., using a broadband antenna) elements of the transceiver system. To keep our study focused, in this paper, we do not discuss any changes to the hardware used by the system and instead concentrate on the signal-processing aspects of the development. 

The remainder of the paper is organized as follows. A generalized transceiver structure that accommodates arbitrary subcarrier placements in Section~\ref{sec:gen_fmtss_trx}. This section also includes details regarding the peak-to-average power ratio (PAPR) and matched-filter responses of contiguous and noncontiguous systems. The details of the receiver signal processing are presented in Section~\ref{sec:rx}. Section~\ref{sec:mod} presents efficient modulation and demodulation structures for both the transmitter and receiver implementations. Finally, Section~\ref{sec:results} presents results comparing the performance of the contiguous and noncontiguous waveforms in both simulated and over-the-air skywave channels. Concluding remarks are given in Section~\ref{sec:conclusion}.

\subsection*{Notations and definitions}
This paper uses both continuous and discrete signals signified with parentheses $(\cdot)$ and brackets $[\cdot]$, respectively. Convolution is represented by the symbol `$\star$'. Vectors and matrices are denoted using bold lowercase and uppercase letters, respectively. The $k$-th element of a vector $\mb x$ is denoted as $x_k$, and the $i$-th column of a matrix $\bm \Theta$ is denoted as $\bm \vartheta_i$. For a time-varying vector $\mb s$, the elements at the $n$-th time index are accessed as $\mb s[n]$. The superscripts $(\cdot)^*$, $(\cdot)^\textup T$, and $(\cdot)^\textup{H}$ indicate complex conjugate, transpose, and Hermitian transpose, respectively. Identity and discrete-Fourier transform (DFT) matrices are represented as $\mb I$ and $\F$, respectively, and DFT matrices are normalized such that $\F^\Hmsp\F=\mb I$. When not clear from the context, we use a subscript to declare the size of these matrices; e.g., $\mb I_M$ is an identity matrix of size $M\times M$. The cardinality of a set $X$ is denoted as $|X|$. Where the distinction is necessary, the subscripts `c' and `nc' indicate quantities with contiguous and noncontiguous subcarrier bands, respectively. 

Several definitions of bandwidth are used to describe the noncontiguous systems. The signal bandwidth $W_\tu s$ is the sum of subcarrier bandwidths occupied by the signal; i.e., in a system with $K$ subcarriers,
\begin{equation}
    W_\tu s = \sum_{k=0}^{K-1} W_\tu{sc},
\end{equation}
where $W_\tu{sc}$ is the subcarrier bandwidth. The channel bandwidth $W_h$ is the width of allowable signal frequencies. The sparsity factor $u$ is a waveform parameter defined as 
\begin{equation}\label{eq:u}
    u=W_h/W_\tu s.
\end{equation}
The sparsity factor $u$ represents how many times wider the channel bandwidth is than the minimum transmission bandwidth, where a value of $u=1$ corresponds to a contiguous waveform. Finally, a noncontiguous signal may not extend over the entire channel bandwidth. The frequency extent $W_\tu e$ of a signal is thus defined as the width from the leftmost to the rightmost occupied frequencies in a noncontiguous signal.

\section{Noncontiguous Transmitter Considerations}\label{sec:gen_fmtss_trx}

\subsection{Signal synthesis}\label{ssec:sig_synth}
\begin{figure}[t]
    \centering
    \includegraphics[width=\columnwidth]{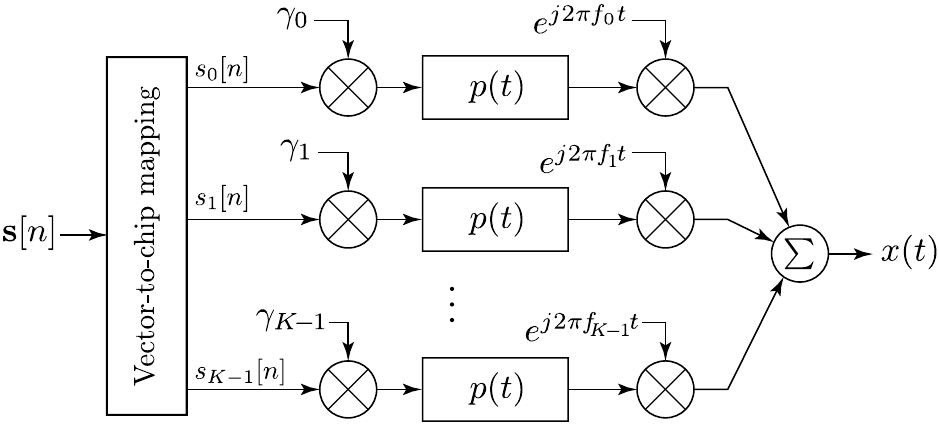}
    \caption{Block diagram of FMT-SS transmitter with multicodes.}
    \label{fig:ncTx}
\end{figure}

Fundamentally, the filtered multitone spread-spectrum (FMT-SS) transmitter synthesizes a waveform $x(t)$ composed of non-overlapping subcarrier bands. A block diagram of the transmitter structure is given in Fig.~\ref{fig:ncTx}. Along the $k$-th subcarrier branch, the symbol chip associated with the $n$-th symbol, $s_k[n]$, is multiplied by the spreading gain $\gamma_k$ before pulse-shaping by the filter $p(t)$. The result is then translated in frequency to the associated subcarrier frequency $f_k$, where $f_k\in F$, and the branches summed together to obtain
\begin{equation}\label{eq:xt}
    x(t) = \sum_n\sum_{k=0}^{K-1}s_k[n]\delta(t-nT_\tu b)\star \gamma_k p(t)e^{j2\pi f_k t},
\end{equation}
where $T_\tu b$ is the symbol interval. In this work, following the previous FMT-SS literature (e.g., \cite{haab_multicode, haab_nmf, wasden, laraway_milcom15, laraway_milcom16}), we let $p(t)$ be a square-root raised cosine filter with an excess bandwidth of $100\%$. Correspondingly, each subcarrier has a bandwidth of $2f_\tu b$, where $f_\tu b=1/T_\tu b$ is the symbol rate.

From \eqref{eq:xt}, it is observed that the elements in $F$ entirely determine whether $x(t)$ has contiguous or noncontiguous subcarriers. For instance, to synthesize a contiguous waveform, the subcarrier center frequencies should belong to the set
\begin{equation}
    F_\tu c = \{ (2k-K+1)f_{\textup b}, ~ k=0,1,\cdots,K-1\}.
\end{equation}
On the other hand, a noncontiguous waveform may consist of any non-overlapping subcarrier frequencies lying within the channel bandwidth $W_h$. In this work, we restrict the subcarrier center frequencies to be odd-integer multiples of the symbol rate. To this end, we first define the superset $\Fnc$ as
\begin{equation}\label{eq:F_superset}
    \Fnc = \{(2k-Ku+1)f_{\textup b}, ~ k=0,1,\cdots,Ku-1\}.
\end{equation}
The set of noncontiguous subcarrier frequencies $F_\tu{nc} \subset \Fnc$ is then formed using one of the following selection criteria.
\begin{itemize}
	\item \textit{Uniformly spaced:} The choices of $k$ in \eqref{eq:F_superset} are equally spaced at the interval $u$.
	\item \textit{Segmented random:} The transmission band is partitioned into $K$ segments, and within each segment, a subcarrier is placed randomly at one of $u$ different positions.
	\item \textit{Random:} The choices of $k$ in \eqref{eq:F_superset} are $K$ unique random integers lying in the interval $0$ to $Ku-1$.
\end{itemize}

\subsection{Data encoding and packet structure}\label{ssec:pkt_struct}
Each transmitted packet has two distinct sections: a preamble and a payload. Each packet begins with a preamble which contains a periodic symbol sequence used for timing acquisition and training in the receiver. The payload section follows the preamble, and it contains the encoded data as well as some pilot symbols to aid channel acquisition and tracking throughout the packet. 

Data in the payload section is encoded using multicodes. As opposed to traditional FMT-SS data encoding in which a single symbol is repeated across each subcarrier branch \cite{Kondo1996}, multicoding applies chip vectors across the subcarrier branches. With multicoding, higher-order modulations can be used without adversely affecting the PAPR \cite{haab_multicode}.

The multicode symbol set is denoted as $\bm \Theta \in \mathbb{C}^{K\times M_\tu d}$, the $i$-th column of which being $\bm \vartheta_i$ and the set containing $M_\tu d$ unique symbols. The elements of $\bm \vartheta_i$ are referred to as symbol chips. The columns of $\bm \Theta$ are typically selected to be orthogonal; e.g., $\bm \Theta$ may be formed by selecting $M_\textup d$ columns of the DFT matrix $\F_{\mspace{-3mu}K}$ \cite{haab_multicode}. The $n$-th data symbol is selected by mapping a group of $\log_2 M_\tu d$ bits to the symbol vector $\bm \vartheta_i$, $0\leq i \leq M_\textup d-1$, and storing the result as the symbol vector $\mb d[n]=\bm \vartheta_i$.

To aid channel estimation, pilot symbols are occasionally transmitted along each subcarrier during the payload. A total of $N_\tu p$ pilot symbols are periodically inserted between the $N_\tu d$ data symbols, leading to the payload symbol matrix 
\begin{multline}\label{eq:payload_structure}
	\mb S = [\mb p_0, \underbrace{\mb d[0], \ldots, \mb d[(\intvl-1)-1]}_{\intvl-1\textup{ elements}},\\
 \mb p_1, \underbrace{\mb d[\intvl-1], \ldots, \mb d[2(\intvl-1)-1]}_{\intvl-1\textup{ elements}}, \mb p_2, \ldots],
\end{multline}
of size $K\times (N_{\tu d}+N_{\tu p})$. Here, $\mb p_l$ is the $l$-th pilot chip vector. At symbol index $n$, the transmitter applies the $n$-th column of $\mb S$ as the input $\mb s[n]$ in to the filterbank shown in Fig.~\ref{fig:ncTx}. The vector $\mb s[n]$ may correspond to either a data symbol or a pilot.

Finally, only data symbols are encoded using multicodes. The training symbols, being part of either the preamble or the inserted pilot sequence, apply the same phase to all the subcarrier branches. That is, the training symbols have the form $\mb z_l=z_l\mb{1}_K$, where $\mb 1_K$ is a vector of $K$ ones and $z_l$ is the $l$-th training symbol.  

\subsection{Peak-to-average power ratio}\label{ssec:papr}
Spreading gains, denoted by $\gamma_k$ in \eqref{eq:xt}, are unity-magnitude scalars applied by the transmitter to reduce the PAPR of the synthesized FMT-SS waveform. These spreading gains are chosen by minimizing the crest factor of the multitone signal
\begin{equation}
    \beta(t) = \sum_{k=0}^{K-1}\gamma_k e^{j2\pi f_k t}.
\end{equation}
This minimization is solved iteratively using the Friese algorithm detailed in \cite{friese}. In \cite{laraway_milcom15}, it is shown that minimizing the crest factor of $\beta(t)$ in turn reduces the PAPR of the synthesized FMT-SS signal $x(t)$. 

To characterize the PAPR behavior of the noncontiguous waveform, we compare NC-FMT-SS signals generated following each of the subcarrier placement methods. Specifically, we generate $500$ noncontiguous signals following each subcarrier placement method detailed in Section~\ref{ssec:sig_synth}. For both of the random subcarrier placements, the subcarrier frequencies $F_\tu{nc}$ are reselected at each of the $500$ trials. For each trial, the spreading gains are optimized using the Friese algorithm, \cite{friese}, to minimize the PAPR of the waveform. 

\begin{figure}[t]
	\centering
	\includegraphics[width=0.9\columnwidth]{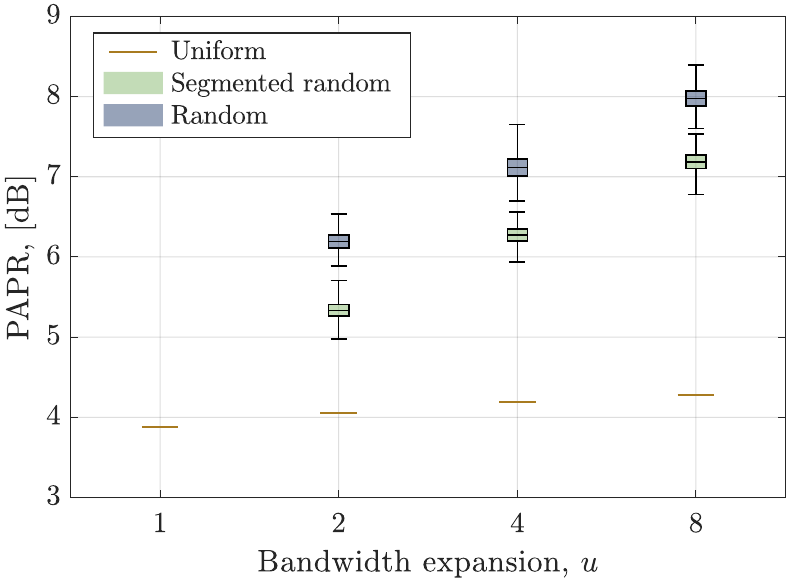}
	\caption{PAPRs for different NC-FMT-SS waveforms with $K=32$ subcarriers for uniform, segmented random, and random subcarrier placements. The colored regions of the boxes extend from the lower to the upper quartiles.}
	\label{fig:papr}
\end{figure}

Fig.~\ref{fig:papr} shows the statistics of the resulting PAPRs. We observe that a uniform subcarrier placement has both the lowest PAPR and the least variance because of the deterministic subcarrier placement. Between the two random methods, the segmented-random placement method offers a lower PAPR. Furthermore, with each doubling of $u$, the PAPR grows by approximately $0.9$~dB for the random placements, but only $0.2$~dB for the uniform placement.

\subsection{Timing acquisition}\label{ssec:timing}
One of the roles of the preamble sequence is timing acquisition. The receiver determines the timing phase by selecting the peak of the correlation between the received preamble sequence and the transmitted preamble known to the receiver. 

As discussed in Section~\ref{ssec:pkt_struct}, each preamble symbol applies the same phase to each subcarrier. We can thus factor \eqref{eq:xt} over the preamble portion of payload $x_\tu{pre}(t)$ as
\begin{equation}\label{eq:x_pret}
    x_\tu{pre}(t) = z(t)\star\left(\sum_{k=0}^{K-1} \gamma_kp(t)e^{j2\pi f_k t}\right),
\end{equation}
where $z(t)=\sum_n z[n]\delta(t-nT_\tu b)$ is the preamble symbol sequence. The summation over $k$ in \eqref{eq:x_pret} represents the impulse response of the transmitter filter bank which is defined as
\begin{equation}\label{eq:gt}
    g(t) = \sum_{k=0}^{K-1} \gamma_kp(t)e^{j2\pi f_k t}.
\end{equation}
Letting $x_\tu{pre}(t)$ pass through a channel with impulse response $h(t)$ and then matched filtering by $g^*(-t)$, we get
\begin{equation}
    y_\tu{pre}(t) =  z(t) \star h(t) \star \eta(t),
\end{equation}
where 
\begin{equation}\label{eq:eta_general}
	\eta(t)=g(t)\star g^*(-t)
\end{equation}
is the overall matched filter response of the transmitter and receiver. The signal $y_\tu{pre}(t)$ is then correlated with the known preamble sequence, yielding
\begin{equation}\label{eq:phi_yz}
    \phi_{yz}(\tau) = \left(\phi_{zz} \star h \star \eta\right)(\tau),
\end{equation}
where $\phi_{zz}(\tau)$ is the autocorrelation of of the transmitted symbol sequence, and $\phi_{yz}(\tau)$ is the cross-correlation between the transmitted and received preamble sequences. The timing phase $\tau_0$ is chosen as the value $\tau$ that maximizes $|\phi_{yz}(\tau)|$. 

To have an unambiguous timing phase estimate, the peak of $\phi_{yz}(\tau)$ should be distinct. For this reason, the preamble symbol sequence $z[n]$ is often chosen such that the autocorrelation of $z[n]$ is $\delta[n]$ (e.g., a Zadoff-Chu sequence). The resulting cross-correlation $\phi_{yz}(\tau)$ thus becomes the convolution of the channel impulse response with the matched filter, meaning the distinctiveness of the correlation peak is determined by the response $\eta(t)$.

\begin{figure}[t]
	\centering
	\includegraphics[width=0.965\columnwidth]{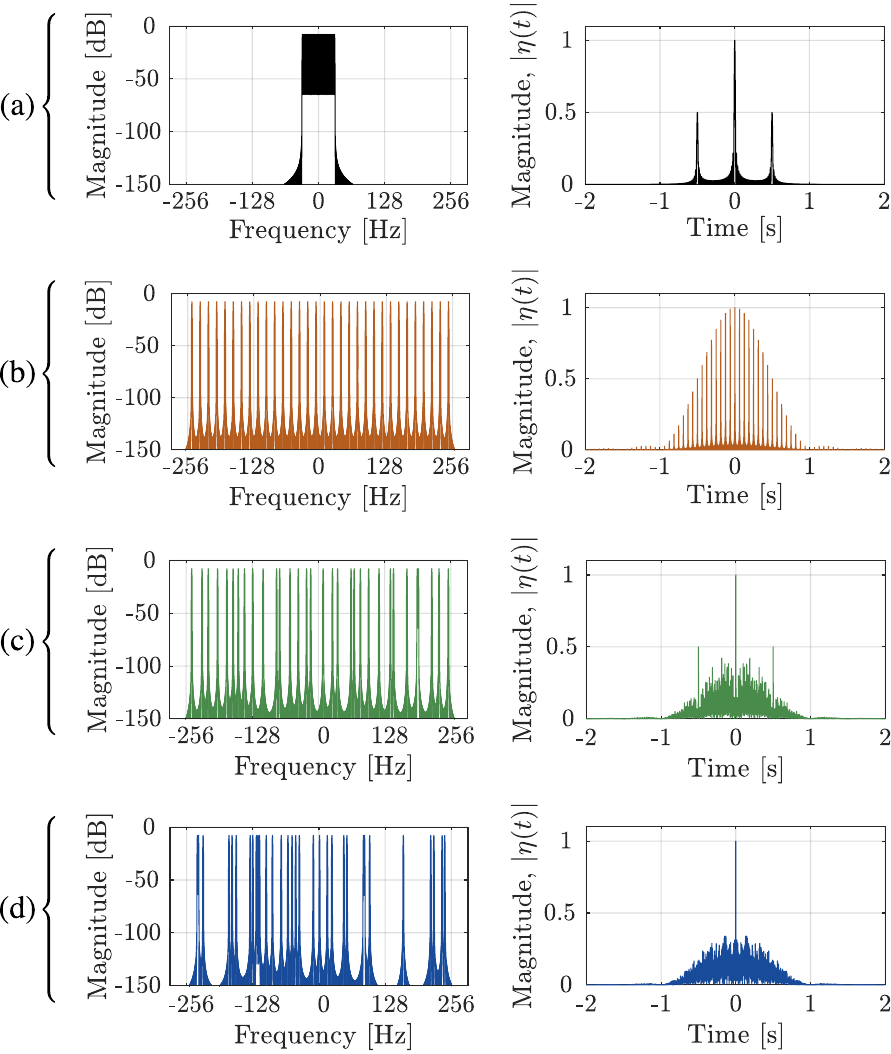}
	\caption{Examples of the power-spectral density (left column) and impulse responses (right column) for: (a) C-FMT-SS; (b) uniform NC-FMT-SS; (c) segmented-random NC-FMT-SS;  (d) random NC-FMT-SS. For the NC-FMT-SS systems, $u=8$. All systems have $K=32$~subcarriers and $f_\tu b = 1$~sym/s.}
	\label{fig:psfs}
\end{figure}

Fig.~\ref{fig:psfs} shows the distinctiveness of the peaks in $\eta(t)$ for both $\eta_\tu c(t)$ as well as $\eta_\tu{nc}(t)$ for each of the noncontiguous subcarrier placement methods. The left and right columns display the power-spectral densities and magnitude responses, respectively, for each subcarrier placement method. Note that the magnitude response $\eta_\tu c(t)$ has only three significant peaks and a clearly resolvable central peak, allowing for straightforward timing recovery. Likewise, both the random and the segmented random placements of NC-FMT-SS subcarriers have responses with a distinct central peak. In contrast, the uniform subcarrier placement has an impulse response with several large-magnitude peaks surrounding its center which may increase the probability of selecting an incorrect timing phase. 

\begin{figure*}[h]
	\centering
	\includegraphics[width=0.8\textwidth]{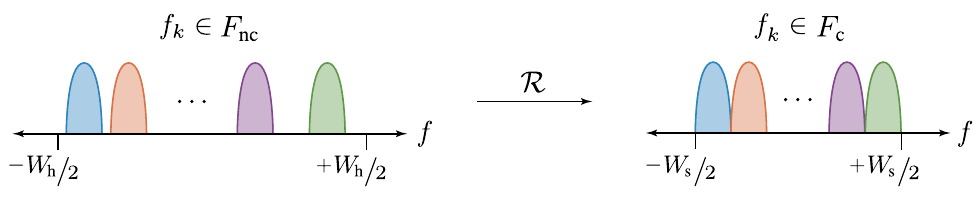}
	\caption{A depiction of $\R$ acting on the subcarrier bands of a noncontiguous signal. Through $\R$, the channel bandwidth $W_h$ is reduced to $W_\textup s$, where $W_\textup s=W_h/u$.}
	\label{fig:Rxfm}
\end{figure*}

\section{Receiver architecture}\label{sec:rx}
Whereas the FMT-SS transmitter structure readily adapts to use with noncontiguous subcarriers, a corresponding receiver implementation is not so straightforward. There are two primary reasons for this: i), the large bandwidth of the transmitted signal requires high sampling rates and, correspondingly, significantly more computations per unit data; and ii), the lack of signal content between subcarriers means the channel impulse response is insufficiently sampled across frequency, leading to a poorly-conditioned channel estimator. In this section, we address both of these points.

\subsection{Subcarrier remapping}\label{ssec:remap}
Due to the large channel bandwidth, noncontiguous signals must be sampled at high rates. However, much of the sampled data is extraneous to recovering the transmitted signal because of its sparsity in the frequency domain. To reduce complexity, the receiver should process only those samples lying in the same frequency subspace as the transmitted signal. To this end, we introduce a pre-processing step in which the noncontiguous center frequencies of the received subcarriers are remapped to their corresponding contiguous positions, removing the nulls between subcarriers in the frequency domain. This procedure is illustrated in Fig.~\ref{fig:Rxfm}. This subcarrier remapping reduces the frequency extent of the signal, thereby reducing the required Nyquist sampling rate by up to a factor of $u$. This step is performed at the front of the receiver, allowing subsequent operations (e.g., matched filtering) to be performed on the transformed, contiguous domain signal.

We denote the remapping operator as $\R$, where the expression $\psi'(t)=\R\{\psi(t)\}$ remaps the subcarriers in the noncontiguous signal $\psi(t)$ to their contiguous positions, yielding the signal $\psi'(t)$. Note that $(\cdot)'$ indicates the signal has been transformed by $\R$. Quantities marked with $(\cdot)'$ differ from those with a subscript `c' in that they have been transformed to the contiguous domain and were not originally formed from the contiguous set $F_\tu c$. 

To determine the impact of the subcarrier remapping on the received signal, we analyze the impulse response $\psi'(t)$ of the system between the input of the transmitter to the input of the receiver matched filter. In the absence of noise, this can be expressed as
\begin{align}\label{eq:psid_unsimple}
    \psi'(t) &= \R\{g_\tu{nc}(t)\star h(t)\},
\end{align}
where $g_\tu{nc}(t)$ is the overall pulse-shaping filter of the transmitter defined in \eqref{eq:gt}, and $h(t)$ is the channel impulse response. The remapping operation $\R$ applied to the received signal $\psi(t)$ comprises three steps: i), taking the Fourier transform of the input; ii), isolating and shifting each subcarrier to its associated contiguous center frequency; and iii), returning the signal to the time domain via the inverse Fourier transform. We thus proceed by applying each of these steps individually. 

First, applying the Fourier transform to $g_\tu{nc}(t)\star h(t)$ provides
\begin{equation}
    \Psi(f) = \left(\sum_k \gamma_k P(f-f_k) \right)H(f),
\end{equation}
where $f_k\in F_\tu{nc}$.

Next, the $k$th subcarrier band in $\Psi(f)$ may be isolated as
\begin{align}\label{eq:Psi_k}
    \Psi_k(f) &= \Pi\left(\frac{f-f_k}{2f_\tu{b}}\right)\Psi(f)\nonumber\\
    &=\gamma_k\Pi\left(\frac{f-f_k}{2f_\tu{b}}\right)P(f-f_k)H(f).
\end{align}
Here, $\Pi(f)$ is the rectangular function centered at zero and having a base of unity. The subcarriers are then shifted to their associated contiguous positions, yielding
\begin{equation}\label{eq:Psik_prime_2}
    \Psi'_k(f) = \gamma_k\Pi\left(\frac{f-f'_k}{2f_\tu b}\right)P(f-f'_k)H(f+f_k-f'_k)
\end{equation}
where $f'_k\in F_\tu c$. 

The overall response considering all subcarriers is then found by summing the individual responses in \eqref{eq:Psik_prime_2}; i.e., 
\begin{align}\label{eq:Psi_prime}
    \Psi'(f)    &= \sum_k \Psi'_k(f) \nonumber \\
                &= \sum_{k_1} \gamma_k P(f-f'_{k_1})\Pi\left(\frac{f-f'_{k_1}}{2f_\tu b}\right) \nonumber \\
                &\mspace{60mu}\times \sum_{k_2} \Pi\left(\frac{f-f'_{k_2}}{2f_\tu b}\right)H(f+f_{k_2}-f'_{k_2}),
\end{align}
where we note the summation can be separated into two terms since the cross terms (i.e., where $k_1\neq k_2$) equal zero and $\Pi(f)\times\Pi(f)=\Pi(f)$. We further note that the rectangular window can be removed in the summation over $k_1$ since $P(f)\times \Pi(f/2f_\tu b) \approx P(f)$. Considering this point, applying the inverse Fourier transform to \eqref{eq:Psi_prime}, and simplifying, we obtain
\begin{equation}\label{eq:psi_prime}
    \psi'(t) = g_\tu c(t) \star h'(t),
\end{equation}
where $g_\tu c(t)$ is the pulse-shaping filter formed using contiguous subcarrier bands, and $h'(t)$ is the equivalent channel impulse response. Explicitly, $h'(t)$ equals 
\begin{equation}\label{eq:h_prime}
    h'(t) = \sum_{k}\left(\sinc(2f_\tu b t)\star h(t)e^{-j2\pi f_k t}\right)e^{j2\pi f'_k t},
\end{equation}
where $\sinc(x)=\tfrac{\sin \pi x}{\pi x}$. Note that $h'(t)$ is equivalent to applying $\R$ directly to the channel impulse response (i.e., $h'(t)=\R\{h(t)\}$).

Comparing the result in \eqref{eq:psi_prime} to \eqref{eq:psid_unsimple}, the remapping transformation converts the received noncontiguous signal into a contiguous signal convolved with the effective channel response $h'(t)$. Hence, subsequent receiver operations can be directly performed on the remapped signal no differently than if it had it been transmitted contiguously. This permits the application of previously established receiver algorithms (e.g., normalized matched filtering (NMF), \cite{haab_nmf}, and multicode symbol detection, \cite{haab_multicode}) given an estimate of the remapped channel $h'(t)$.

\begin{figure*}[t]
	\centering
	\includegraphics[width=0.98\textwidth]{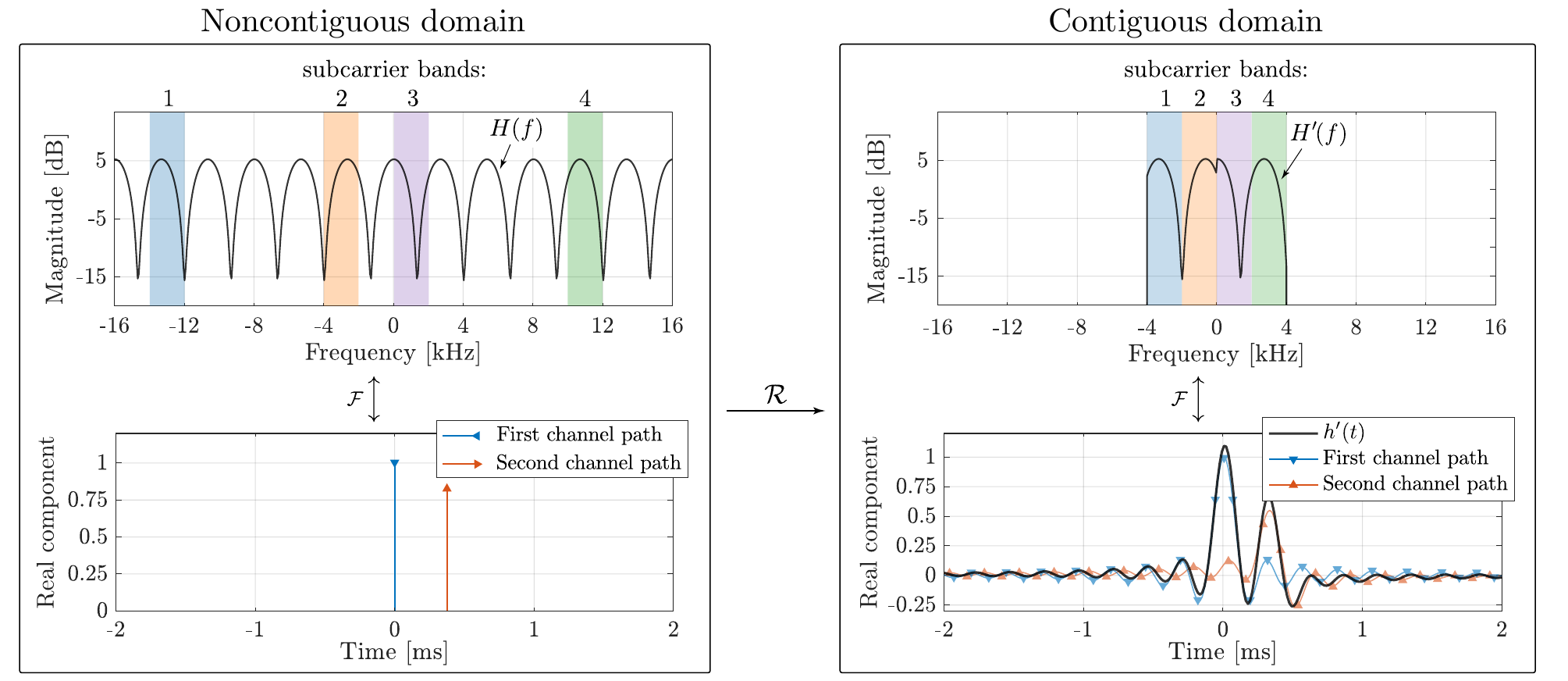}
	 \caption{A frequency and time view of a two-path channel before and after transformation by $\R$ in a system with $4$ subcarriers. The symbol {$\mathcal{F}\big\updownarrow$} indicates a Fourier transform pair. Because each subcarrier band is isolated by $\R$, $H'(f)$ is the concatenation of the channel frequency response over each noncontiguous subcarrier band.}
	\label{fig:Tchan}
\end{figure*}

Considering \eqref{eq:h_prime}, the remapped channel will be spread in the time-domain through $\R$ due to the introduction of sinc pulses. Fig.~\ref{fig:Tchan} illustrates this effect by comparing a simple two-path channel $h(t)$ to its counterpart $h'(t)$ after transformation by $\R$. Note that the original channel $h(t)$ is a continuous function of frequency. In contrast, the remapped channel $h'(t)$ has a piecewise frequency response which concatenates the individual narrow responses over each subcarrier, yielding the response $H'(f)$. The effect of this concatenation is time-domain spreading, giving rise to the sinc terms in \eqref{eq:h_prime} and demonstrated by $h'(t)$ in Fig.~\ref{fig:Tchan}.

\subsection{Discrete-time receiver processing}
Continuous time analysis was used in the previous section to introduce the remapping operator $\R$. Here, to facilitate the development of practical receiver algorithms, we introduce the sampled-equivalent of $\R$ using matrices and vectors. 

Note that the primary objective of the remapping transformation is to simplify receiver processing by removing extraneous frequency content from the received signal. After remapping, the signal comprises a contiguous band of subcarriers centered around $0$~Hz. Through this process, the Nyquist rate of the signal is reduced, thus allowing the remapped time-domain signal to be downsampled. However, the signal can equivalently be downsampled in the frequency domain: instead of shifting the subcarriers to their associated contiguous positions, the FFT bins in the nulls between subcarriers can be discarded such that the resulting frequency-domain vector contains a contiguous band of subcarriers. The matrix equivalent of $\R$ is then concisely represented by $\mb R$,
\begin{equation}\label{eq:R_defn}
    \mb{R} = \F_{M'}^\Hmsp \Ii \F_M,
\end{equation}
where $\F_M$ is a DFT matrix of size $M$, and $\Ii$ is an $M'\times M$ selection matrix, $M'=M/u$. The matrix $\Ii$ is formed by selecting the rows from $\mb I_M$ corresponding to the frequency bins of the subcarriers in the transmitted signal. 

Consider the received noncontiguous vector $\mb y$ of length $M$. This vector is remapped by the product $\mb y'= \mb{Ry}$, where $\mb y'$ is of length $M'$. Following \eqref{eq:R_defn}, the DFT matrix $\F_M$ converts the vector $\mb y$ to the frequency domain, yielding $\mb y_\tu{f}$. The matrix $\Ii$ keeps only the bins of $\mb y_\tu{f}$ corresponding to the positions of the noncontiguous subcarriers, where the resulting vector is finally returned to the time domain by the matrix $\F_{M'}^\Hmsp$. Note that multiplying a contiguous-domain vector by $\mb R^\Herm$ instead yields a vector with noncontiguous subbands.

\subsection{Channel acquisition}\label{ssec:chan_acq}
The channel estimation architecture presented here is based on the two-stage approach for HF channels introduced in \cite{milcom23}. In the first stage, the power-delay profile (PDP) of the channel is estimated using the inserted pilot symbols. The second stage of the estimator then uses the sparsity and power information from the PDP to estimate the nonzero taps in the channel impulse response vector. In \cite{milcom23}, this method was limited to contiguous signals; here, we expand the approach for use with noncontiguous signals.

In the recevier, there are two difference channel responses present: the actual channel impulse response $h(t)$, and the remapped channel impulse response $h'(t)$. Since symbol detection is performed on the remapped signal, the receiver ultimately needs an estimate of $h'(t)$. However, recall both from \eqref{eq:h_prime} and Fig.~\ref{fig:Tchan} that $\R$ spreads the channel impulse response in time. That is, through $\R$, a channel $h(t)$ with a finite support is transformed to $h'(t)$ with an infinite support. The spreading of $h'(t)$ decays slowly following a $\sinc$ pulse whose bandwidth is $2f_\textup b$. This slow decay poses the following dichotomy: an estimate of $h'(t)$ should be stored in a vector long enough to include most of the energy in the channel response, but short enough to exclude the low-valued channel taps that are dominated by noise and degrade the estimate \cite{mst}. Estimating $h'(t)$ is thus a balance between energy recovery and estimation accuracy.

\begin{figure}[t]
	\centering
	\includegraphics[width=\columnwidth]{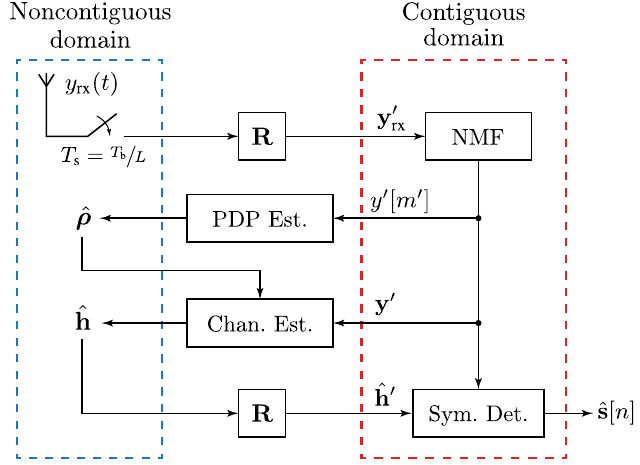}
	\caption{The receiver block diagram sorted by domain. The signal $y_\textup{rx}(t)$ is sampled every $T_\textup s=1/f_\textup s=T_\tu b/L$ seconds. The PDP estimate $\hat{\bm \rho}$ is input to the channel estimator to produce the channel estimate $\hat{\mb h}$. This vector is then transformed to the contiguous domain and input to the RAKE symbol detector.}
	\label{fig:ncSymDet}
\end{figure}

Alternatively, the actual channel impulse response $h(t)$ has finite support and can be represented completely in a finite-length vector. Additionally, an estimate $\hat h(t)$ in the noncontiguous domain is readily transformed to the contiguous domain by $\hat h'(t)=\R\{\hat h(t)\}$. For these reasons, we proceed by developing a noncontiguous channel estimator that provides $\hat{\mb h}$. The resulting estimate is then converted for use by the symbol detector by the product $\hat{\mb h}'=\mb{R}\hat{\mb h}$. A block diagram of the proposed structure is presented in Fig.~\ref{fig:ncSymDet}.

\subsubsection{PDP estimator}\label{ssec:pdpEst}
The PDP is estimated using the pilot symbols inserted throughout the payload section of the packet. Following Fig.~\ref{fig:ncSymDet}, the PDP estimator is given signal samples after remapping by $\mb R$ and matched filtering by the NMF. Next, we form the matrix
\begin{equation}
	\mb Y' = [\mb y'_0~ \mb y'_1~ \ldots ~ \mb y'_{N_{\textup p}-1}],
\end{equation}
where $\mb y'_l$ has length $L'$ and contains the received signal samples centered about the $l$-th received pilot symbol. The length $L'$ of each received pilot-symbol vector $\mb y'_l$ should be long enough to contain the entire channel impulse response $h'(t)$. However, strictly speaking, the spreading introduced by sinc pulses in \eqref{eq:h_prime} extends the duration of the channel response to infinity. Nevertheless, because the sinc function decays proportionally to the distance from its center, so long as the length $L'$ extends for a sufficient duration on either side of the channel impulse response, $\mb y'_l$ will capture a majority of the channel energy. Here, we have empirically determined $6$ symbol intervals to be a sufficient duration.

A delay-time representation of the transformed channel impulse response is then obtained as 
\begin{equation}\label{eqn:cdt}
	\hat{\mb H}'_{\eta} = \mb{Y'P}^*\mspace{-5mu},
\end{equation}
where $\mb P$ is a diagonal matrix comprising the pilot symbols $p_0, p_1, \ldots, p_{N_\tu p-1}$. Multiplying $\mb Y'$ by $\mb P^*$ removes the phase of each pilot symbol, such that each transmitted pilot symbol becomes an impulse. The columns of $\hat{\mb H}'_{\eta}$ are thus the cascade of the impulse response of the transmitter, the channel, $\mb R$, and the NMF. Recalling \eqref{eq:gt} and \eqref{eq:psi_prime}, this is equivalent to the impulse response of the contiguous-band matched filter $\eta_\tu{c}(t)$ convolved with the transformed channel response $h'(t)$.

Next, we apply the inverse remapping operation, yielding
\begin{equation}
	\hat{\mb H}_{\eta} = \mb R^\Herm \hat{\mb H}'_{\eta}.
\end{equation}
Each column in $\hat{\mb H}_{\eta}$ is thus the noncontiguous matched filter response $\eta_\tu{nc}(t)$ convolved with the original channel impulse response $h(t)$, and each row of $\hat{\mb H}_{\eta}$ contains the time variation of each channel tap. The columns and rows of $\hat{\mb H}_{\eta}$ thus capture the channel variation along the delay and time dimensions, respectively. Applying the DFT to each row and squaring the magnitude of the resulting matrix thus provides the Doppler spectrum of each channel tap. The resulting matrix is called the scattering function and is denoted as $\Seta$. Considering that the columns of $\hat{\mb H}_{\eta}$ are the convolution of $h(t)$ and $\eta_\tu{nc}(t)$, one finds that the columns of $\Seta$ thus contain the power-delay profile of the channel impulse response filtered by $|\eta_\textup{nc}(t)|^2$. 

An intermediate PDP $\hat{\bm \rho}_\eta$ is then obtained by performing a weighted sum across the rows of $\Seta$. The weights in the summation should be chosen to approximately match the channel response's anticipated Doppler profile \cite{milcom23}. Following the preceding discussion, we can then relate the channel PDP $\bm \rho$ to the intermediate PDP $\hat{\bm \rho}_{\eta}$ as 
\begin{equation}
	\hat{\bm \rho}_{\eta} = \mb N {\bm \rho} + \bm \nu,
\end{equation}
where $\mb N$ is a circulant matrix that implements convolution with $|\eta_\tu{nc}(t)|^2$ and $\bm \nu$ is a noise plus interference term.

Finally, because skywave HF channels are sparse in delay \cite{johnson2012}, the channel PDP $\bm \rho$ will be sparse as well. To promote a sparse PDP estimate $\hat{\bm \rho}$, we cast the problem as a basis pursuit denoising (BPDN) optimization, \cite{theodoridis}, such that
\begin{equation}\label{eqn:ncOpt}
	\rhoc = \argmin_{\rhoc}~\lVert\rhoc\rVert_{\ell_1}\quad\textup{s.t.}\quad\lVert{\hat{\bm \rho}}_{\eta} - \mb N\rhoc\rVert_{\ell_2} < \varepsilon,
\end{equation}
where $\varepsilon$ is a threshold value that depends on the noise level.

There are many algorithms that can be used to solve for $\hat{\bm \rho}$ in \eqref{eqn:ncOpt}. In \cite{milcom23}, \eqref{eqn:ncOpt} was solved using the orthogonal matching pursuit algorithm (OMP) \cite{omp}. Here, we note that the underlying structure of the PDP in HF channels is block sparse. That is, the nonzero values of channel delays occur in clusters referred to as modes in the context of HF channels and known as blocks in the context of sparse approximation. Correspondingly, we use the weighted double-backtracking matching-pursuit (WDBMP) algorithm \cite{wdbmp}. This algorithm is used for its low complexity and performance with unknown block sizes and boundaries. 

Regarding implementation, instead of estimating the signal-to-noise ratio (SNR) to determine the value of $\varepsilon$ in \eqref{eqn:ncOpt}, we instead terminate the WDBMP loop based on the difference of successive PDP estimates. Denoting the PDP estimate at the $i$-th iteration of the WDBMP loop as $\hat{\bm \rho}_i$, the normalized mean-squared difference $\kappa_i$ between the current $i$-th and previous $(i-1)$-th PDP estimates is measured as
\begin{equation}
	\kappa_{i} = \frac{(\hat{\bm \rho}_i - \hat{\bm \rho}_{i-1})^\Herm(\hat{\bm \rho}_i - \hat{\bm \rho}_{i-1})}{(\hat{\bm \rho}_{i})^\Herm(\hat{\bm \rho}_{i})}.
\end{equation}
In this work, the loop terminates when $\kappa_i<10^{-3}$.

The estimate $\rhoc$ is lastly convolved with a Gaussian pulse to smear the estimated support of the PDP. This smearing increases the number of falsely selected taps, but also reduces the number of nonzero taps in $\bm \rho$ that were missed in $\hat{\bm \rho}$. The smoothing filter used herein has a total duration of $100~\mu$s and a $2\sigma$ width of $22~\mu$s.

\subsubsection{Channel estimation}\label{ssec:chanEst}
After the receiver estimates the PDP, the channel impulse response is estimated at various points across the payload. An initial channel estimate is obtained using the periodic preamble sequence described in Section~\ref{ssec:pkt_struct} which begins each transmission. This initial estimate is subsequently updated throughout the payload section of the packet using the inserted pilot symbols. For brevity, here, we develop the channel estimator using only the preamble sequence; details regarding channel tracking are given in \cite{milcom23}. 

A single period of the received, contiguous domain (i.e., remapped) preamble is denoted as ${\mb y}'$. The vector ${\mb y}'$ has length $M'=2KZ$, where the system has $K$ subcarriers, and there are $Z$ symbols in each period of the preamble. The preamble ${\mb y}'$ may be expressed as 
\begin{equation}\label{eqn:sigrx}
	\ypre = \mb{G'_\textup{NMF}R}({\mb y_\tu{rx}} + \mb v),
\end{equation}	
where $\mb G'_{\textup{NMF}}$ is a circulant matrix implementing the contiguous-domain NMF at the receiver input, $\mb y_\tu{rx}$ is one period of the received noncontiguous domain preamble with length $M=u\cdot2KZ$, and $\mb v$ represents the noise plus interference.

Because the preamble symbol sequence is periodic, we can express $\mb y_\tu{rx}$ as a product of circulant matrices \cite{haab_nmf}; viz., 
\begin{equation}\label{eq:yrx_exp}
    \mb y_\tu{rx}=\mb {GZh},
\end{equation}
where $\mb G$ and $\mb Z$ are circulant matrices implementing the convolution of the noncontiguous pulse-shaping filter at the transmitter and expanded preamble symbols, respectively, and $\mb h$ is the channel impulse response. Replacing $\mb y_\tu{rx}$ with \eqref{eq:yrx_exp} and substituting $\mb R$ with its definition in \eqref{eq:R_defn}, \eqref{eqn:sigrx} can be written as
\begin{equation}\label{eqn:yPreDir}
	\mb y' = \mb G'_\textup{NMF}\F_{M'}^\Hmsp\Ii\F_M \mb{GZh} + \bar{\mb v}',
\end{equation}
where
\begin{equation}
    \bar{\mb v}'=\mb G'_\tu{NMF}\mb{Rv}
\end{equation}
is the filtered noise plus interference lying in the signal subspace.

In \cite{grayCirc}, it is shown that a circulant matrix $\mb A$ can be decomposed as
\begin{equation}\label{eqn:diag}
	\mb A = \F^\Hmsp\!\subf{\mb A} \F, 
\end{equation}
where $\subf{\mb A}$ is a diagonal matrix whose nonzero elements are the DFT of the first column of $\mb A$. Applying \eqref{eqn:diag} to $\mb G'_\textup{NMF}$, $\mb G$, and $\mb Z$ in \eqref{eqn:yPreDir}, $\mb y'$ can be written as
\begin{equation}\label{eqn:yPrimeInt}
	\mb y' = \F_{M'}^\Hmsp\mb G'_\textup{NMF, f}\mspace{2mu}\Ii\subf{\mb G}\subf{\mb Z}\F_M\mb h + \bar{\mb v}'.
\end{equation}
Note that for any diagonal matrix $\subf{\mb A}$, $\Ii \subf{\mb A}=\subf{\mb A'}\mspace{1mu}\Ii$, where $\subf{\mb A}$ has noncontiguous bands, and $\subf{\mb A'}$ contains only the respective contiguous bands. Additionally, note that $\mb G'_\textup{NMF, f}=\subf{\mb Q'}\subf{\mb G'}^*$, where $\subf{\mb Q'}$ is a diagonal matrix containing the normalization coefficients of the NMF in the frequency domain \cite{haab_nmf}. Combining these two points, \eqref{eqn:yPrimeInt} may be rearranged as
\begin{align}
	\mb y' 	&=\F_{M'}^\Hmsp\subf{\mb Q'}\subf{\mb G'}^*\subf{\mb G'}\subf{\mb Z'}\mspace{2mu}\Ii\F_M\mb h + \bar{\mb v}'.
\end{align}

Recall that the support of the vector $\mb h$ is known from the PDP estimate $\rhoc$. We thus proceed by estimating only the nonzero coefficients of $\mb h$. To this end, we replace $\mb h$ by $\subp{\mb h}$, where $\subp{\mb h}$ is a pruned vector that retains only the elements corresponding to the support of $\rhoc$. The columns of $\F_M$ are similarly pruned to the support of $\rhoc$. This leads to
\begin{equation}
	\ypre 	= \F_{M'}^\Hmsp \Xpref \subp{\mb h} + \bar{\mb v}',
\end{equation}
where 
\begin{equation}\label{eq:Xpref}
    \Xpref=\mspace{2mu}\subf{\mb Q'}\subf{\mb G'}^* \subf{\mb G'}\mspace{2mu}\subf{\mb Z'}\mspace{2mu}\Ii\F_{M\mspace{-4mu},\mspace{2mu}\textup{p}}
\end{equation}
is the filtered preamble sequence, and the subscript `p' in the matrix $\F_{M\mspace{-4mu},\mspace{2mu}\textup{p}}$ denotes the pruned version of $\F_M$. 

The MMSE estimate of the channel impulse response is thus be obtained as \cite{kay}
\begin{equation}\label{eqn:chat0}
	\subp{\hat{\mb h}} = \big(\hat{\bm{\Sigma}}_{\mb h_\textup{p} \mb h_\textup{p}}^{-1} + \XprefH\sigvdvd^{-1}\Xpref\big)^{-1}\mb \XprefH \sigvdvd^{-1} \ypref,
\end{equation}
where $\hat{\bm{\Sigma}}_{\mb h_\textup{p} \mb h_\textup{p}}$ is the channel-tap covariance matrix whose diagonal is the PDP estimate $\hat{\bm \rho}$, $\sigvdvd$ is the covariance matrix of the noise plus interference vector $\subf{\bar{\mb v}'}$, and $\ypref$ is the scaled Fourier transform of the received preamble $\ypre$ (see \eqref{eq:yscale} below). Finally, the channel impulse response estimate is obtained by letting the nonzero tap positions of $\hat{\mb h}$ equal $\subp{\hat{\mb h}}$. 

To ensure $\hat{\mb h}$ is properly scaled, we define $ \ypref$ as
\begin{equation}\label{eq:yscale}
	 \ypref = \tfrac{1}{\sqrt{u}}\F_M\ypre.
\end{equation}
The factor $1/\sqrt u$ is added to return the signal level to its proper magnitude. Recall the definition of $\mb R$ in \eqref{eq:R_defn} contains the cascade of an IDFT matrix of size $M'$ with a DFT matrix of size $M$. Since these matrices are normalized such that $\F_{M'}^\Hmsp\F_{M'}=\mb I_{M'}$ and $\F_{M}^\Hmsp\F_M=\mb I_M$ where $M=uM'$, the signal after transformation by $\mb R$ is inadvertently amplified by $\sqrt{u}$.

Regarding the noise covariance matrix, the contribution due to noise and interference in $ \ypref$ is $\bar{\mb v}'_{\mspace{-1.5mu}\textup{f}}=\tfrac{1}{\sqrt{u}}\mspace{2mu}\F_{M'}\mb{Q'G'^\Herm R} \mb v$. Recalling \eqref{eq:R_defn} and applying \eqref{eqn:diag}, this can be expanded as
\begin{equation}
	\bar{\mb v}'_{\mspace{-1.5mu}\textup{f}} =\tfrac{1}{\sqrt{u}}\subf{\mb Q'}\subf{\mb G'}^*\Ii\F_{M}\mb v.
\end{equation}
The covariance matrix $\sigvdvd$ may thus be expressed in terms of $\bm \Sigma_{\subf{\mb v}\subf{\mb v}}$ as
\begin{equation}\label{eqn:covPart1}
	\sigvdvd 	= \tfrac{1}{u}\mspace{1mu}\subf{\mb Q'}\subf{\mb G'^*}\Ii\bm{\Sigma_{\subf{\mb v}\subf{\mb v}}}\Ii^{\textup{T}}\subf{\mb G'}\subf{\mb Q'^{*}}.
\end{equation}

\subsubsection{Channel recovery criterion}\label{sssec:chanRecov}
Since the actual channel impulse response $h(t)$ is estimated using a noncontiguous waveform, the channel impulse response is undersampled according to the Nyquist rate (e.g., in Fig.~\ref{fig:Tchan}, while the Nyquist rate of the input signal is $32$~kHz, the transformation $\R$ retains only $8$~kHz of received signal content) \cite{spfsdr}. In other words, estimating $\hat{\mb h}$ involves solving an underdetermined system of equations (see \eqref{eqn:yPreDir}). Nevertheless, the preceding section details a channel estimator that solves this system by exploiting the sparse structure of the underlying signal $\mb h$. This method of solving underdetermined systems is referred to as compressed sensing. Through the compressed sensing framework, we bound the sampling requirements and determine the conditions under which the estimate of $\mb h$ can be reliably obtained.

The channel impulse response $\mb h\in \mathbb{C}^{M}$ is a sparse vector with support $T$. Let $\Omega$ denote the set of frequencies along which $\mb h$ is sampled. Then, 
\begin{equation}
	\Omega \subset \{0, \ldots, M-1\}\cdot f_\textup s /M,
\end{equation}
and $|\Omega|=M'$. The channel vector $\mb h$ can be recovered from the frequency domain samples $\Omega$ when, \cite{candes2006}, 
\begin{equation}\label{eqn:sampBound}
	|T| \leq \tfrac{1}{2}|\Omega|.
\end{equation}

Using \eqref{eqn:sampBound} and recalling that $|\Omega|=M'=2KZ$, recovery of the channel would be possible only when there are at most $KZ$ nonzero elements in the channel impulse response vector. 

As an example, suppose we are using an NC-FMT-SS system with $K=32$~subcarriers, $Z=16$~preamble symbols, $f_\textup b = 1$~kHz, and $u=8$. We transmit over a channel with a total nonzero duration of $\tau$~seconds; hence, the support of $\mb h$ is approximately $\tau f_\textup s$. Following \eqref{eqn:sampBound}, we can recover the channel when
\begin{equation}\label{eqn:tau1}
	\tau f_\textup s \leq \tfrac{1}{2}\cdot 2KZ.
\end{equation}
The Nyquist sampling rate of the noncontiguous signal is $f_\textup s=W_h=2Kuf_\textup b$. Hence, \eqref{eqn:tau1} can be rearranged as
\begin{equation}
	\tau \leq \frac{Z}{2uf_\textup b}.
\end{equation}
For the described NC-FMT-SS system, we can recover $\mb h$ so long as $\tau \leq 1$~ms. 

Note that \eqref{eqn:sampBound} provides an absolute upper limit; practical results suggest that $|T|$ should be no greater than $\tfrac{1}{3}|\Omega|$ to $\tfrac{1}{5}|\Omega|$ \cite{theodoridis}. Furthermore, the bound holds only for \textit{random} subsets $\Omega$. In an NC-FMT-SS system, the selection of frequencies is not fully random and is instead limited to $K$ swaths of continuous bandwidth. Hence, determining exact recovery conditions for the channel response is unclear due to the ambiguity surrounding the proper coefficient on $|\Omega|$. A more thorough study into the theoretical and experimental recoverability requirements of the channel response, as they apply to an NC-FMT-SS system, is necessary and should be addressed in future studies. Nevertheless, the numerical results presented in Section~\ref{sec:results} indicate that the proposed channel estimator delivers acceptable performance in all the cases we have examined, including a number of over-the-air transmissions. 

\section{Efficient transceiver implementation}\label{sec:mod}
The sparsity factor $u$ defined in \eqref{eq:u} determines how much additional bandwidth the noncontiguous waveforms occupy as compared to their contiguous counterparts. Large values of $u$ grant noncontiguous systems more frequency agility, but they also mandate higher sampling rates, longer convolutions, and, hence, increased complexity. 

To prevent the noncontiguous systems from becoming prohibitively complex, the long convolutions can be performed efficiently in the frequency domain using the overlap-save method \cite{proakis_dsp}. Further, because the NC-FMT-SS signal comprises several modulated narrowband subcarriers, the overlap-save convolution step can be combined with the modulation step to efficiently synthesize the transmit signal directly from the input symbols; this approach is termed fast convolution (FC) \cite{renfors2014}. 
 
To simplify the notation for the multirate FC structures, we let the rate of a signal be implied by its time index. In particular, the time index `$n$' refers to a signal sampled at the symbol rate, and `$m$' refers to sampling rate of the synthesized noncontiguous signal. For example, given a continuous-time signal $x(t)$, its discrete-time counterpart $x[n]$ is sampled at the symbol rate, whereas $x[m]$ is sampled at output rate.

\subsection{Transmitter}
\begin{figure}
	\centering
	\includegraphics[width=\columnwidth]{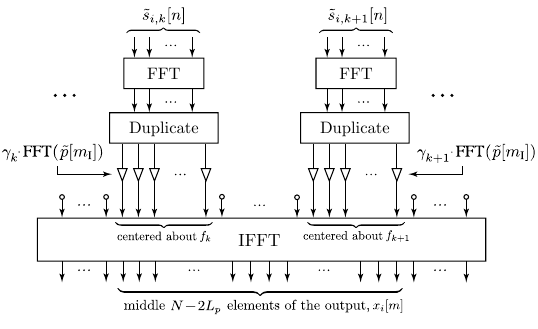}
	\caption{NC-FMT-SS transmitter structure implemented using fast-convolution.}
	\label{fig:fctx}
\end{figure}

Fig.~\ref{fig:fctx} presents a diagram of the proposed transmitter structure. Following the discussion regarding multicodes in the preceding section, we allow each subcarrier to carry different symbol sequences. The sequence of symbols for the $k$-th subcarrier is denoted as $s_k[n]$. By sampling \eqref{eq:xt} at the rate $f_\textup s=Lf_\textup b$, we obtain the modulated output signal as
\begin{equation}\label{eqn:directImp}
	x[m] = \sum_{n} \sum_{k=0}^{K-1} s_k[n]\delta[m - nL]\star \gamma_k p[m]e^{j2\pi (f_k/f_\textup{s}) m},
\end{equation}
where $f_k\in F$, and there are $L=2Ku$ samples per symbol interval.

Following \cite{renfors2014}, when filtering the symbol sequence, only the passband and transition band of the filter $p[m]$ are considered in the FC structure. The filter is thus indexed by $m_\textup I$, where $p[m_\textup{I}]$ is critically sampled, and the subscript `I' asserts the filter operates at an intermediate rate lying between the symbol rate $f_\textup b$ and the sampling rate $f_\textup s$. Thus, including $n$ and $m$, there are a total of three rates in the transmitter which are related to the sampling frequency $f_\textup s$ by 
\begin{equation}
	f_\textup{s} = Kuf_\textup{I} = 2Ku f_\textup b.
\end{equation}
The prototype filter $p[m]$ has a length of $2L_p + 1$ samples; at the intermediate rate, $p[m_\textup I]$ has a length of $2L_{p_\textup I} + 1$, where $L_{p_\textup I}=L_p/Ku$. We section the symbol sequences into groups containing $N_s$ consecutive time domain symbols, where $N_s$ is an integer power of $2$, and each symbol group begins with the last $L_{p_\textup I}/2$ symbols from the preceding section. We denote the $i$-th symbol group for the $k$-th subcarrier as $\tilde s_{i, k}[n]$.

The symbol groups $\tilde s_{i,k}[n]$ are converted to the frequency domain via an FFT, after which their spectra are duplicated (this duplication is equivalent to $2$-fold expansion, thus bringing $\tilde{s}_{i, k}[n]$ to the intermediate rate $f_\textup I$). We then form the filter $\tilde p[m_\textup I]$ by padding $p[m_\textup I]$ with zeros to length $2N_s$. The frequency domain symbol samples are then multiplied by the FFT of $\tilde{p}[m_\textup I]$ and by the spreading coefficients $\gamma_k$ associated with each subcarrier. Next, the results from these operations are shifted to their corresponding subcarrier bands before being brought back into the time domain through an IFFT of size $N=2KuN_s$. The signal at the output of this IFFT operates at the sampling rate $f_\textup s$. By choosing $p[m_\textup I]$ to be a zero phase filter, then following the overlap-save method, the middle $N-2L_p$ samples at the output of the IFFT block are saved; this process is repeated for the remaining symbol blocks \cite{renfors2014}.

\subsection{Receiver}
\begin{figure}
	\centering
	\includegraphics[width=\columnwidth]{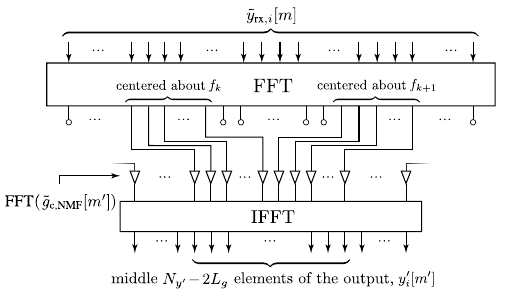}
	\caption{NC-FMT-SS receiver structure implemented using fast-convolution.}
	\label{fig:fcrx}
\end{figure}

Recall that the transformation $\R$ described in Section~\ref{ssec:remap} operates by isolating each received subcarrier, shifting it to its corresponding contiguous position, and then taking the IFFT of the resulting signal. This procedure naturally pairs with FC, such that the remapping and the NMF operations can both be performed as part of the FC structure. The resulting receiver structure is shown in Fig.~\ref{fig:fcrx}.

The receiver operates on the received signal $y_\textup{rx}[m]$ sampled at the rate $f_\textup s$. This signal is sectioned into groups containing $N_y$ samples, where each group begins with the last $L_p$ samples of the preceding section. The $i$-th group is denoted as $\tilde y_{{\textup{rx}}, i}[m]$. 

The $i$-th group is then passed through an FFT, after which the receiver performs the remapping by discarding the output frequency bins not corresponding to the transmitted subcarriers. The dimensionality of the signal group is thus reduced by a factor $u$, such that the resulting vector contains $N_{y'}=N_y/u$ elements and is sampled at the equivalent contiguous rate $f'_\textup{s} = 2Kf_\textup b$. By discarding the unused samples, the remaining samples in the signal become centered about their corresponding contiguous subcarrier frequencies $f'_k$, where $f'_k\in F_\textup c$. 

Next, the receiver convolves the signal with the matched filter. From \eqref{eq:psi_prime}, this is the time-reversed conjugate of the transmitter filter constructed using contiguous subcarrier bands. Here, the received filtering is performed using a normalized matched-filter (NMF), $g_\textup{c,NMF}[m']$, with length $2L_g+1$ samples and $m'$ corresponding to the rate $f'_\textup s$. The construction of this filter is detailed in \cite{sibbett_nmf, haab_nmf}, and is the cascade of a matched-filter $g_{\textup c}^{*}[-m']$ and a normalization filter $q_\textup{c}[m']$ which serves to equalize the power of the subcarrier bands. Note that the filter $g_{\textup c}^{*}[-m']$, by its definition in \eqref{eq:gt}, removes the impact of the spreading gains $\gamma_k$ on each subcarrier.

Mirroring the treatment of $p[m_\textup I]$ in the transmitter, $g_\textup{c,NMF}[m']$ is zero-padded to produce $\tilde g_\textup{c,NMF}[m']$ with length $N_{y'}$. The signal samples are then multiplied by the frequency domain coefficients of $\tilde g_\textup{c,NMF}[m']$, the IFFT of the result is taken, and the first and last $L_g$ samples are discarded to obtain $y'_i[m']$. This process is repeated for each group of samples to produce the filtered and transformed output signal $y'[m']$.

\subsection{Selection of subcarrier frequencies}\label{ssec:selFk}
To use the structures in Figs.~\ref{fig:fctx} and~\ref{fig:fcrx}, certain conditions must be satisfied. In particular, these structures require the central tone of each subcarrier to end on a complete period for each stride of output samples to avoid introducing phase discontinuities in the signal \cite{borgerding}. When $\Fnc$ is defined as in \eqref{eq:F_superset}, this condition is satisfied for any $F\subset\Fnc$.

\section{Performance Results}\label{sec:results}
In this section, we evaluate the performance of the NC-FMT-SS system in various simulated channel conditions as well as in a $300$~km over-the-air (OTA) link between Idaho Falls, Idaho, USA, and Salt Lake City, Utah, USA. 

To generate the simulated results, the wideband HF channel model introduced in \cite{mastrangelo} is used. This model is selected in favor of the more commonly used Watterson model, \cite{watterson}, because it reflects the increased channel-tap resolution observed by wideband waveforms. We select the channel parameters to follow the mid-latitude disturbed (MLD) channel from the ITU recommendation F.1487 \cite{itur_recommendation}. Specifically, the channel has $2$ equal power modes separated by $2$~ms. The RMS delay spread of each mode is $\tau_\textup d=80~\mu$s, and each mode fades with a Gaussian Doppler spectrum whose $2\sigma$ width is $1$~Hz.

Four different configurations of FMT-SS systems are tested using the parameters given in Table~\ref{tab:testParams}. These tests comprise a C-FMT-SS system (corresponding to $u=1$) and three NC-FMT-SS systems with $u$ equaling one of $2$, $4$, or  $8$. Each system has a signal bandwidth of $W_\tu s=~64$~kHz which is comparable to the widest bandwidths used in \cite{hf1363}. To simplify the presentation, none of the systems use forward error correction. Finally, a segmented random subcarrier placement method is used since it offers a compromise between low PAPRs and unambiguous timing acquisition (see Figs.~\ref{fig:papr} and~\ref{fig:psfs}).

To generate the simulated results, for each packet, a new channel realization is generated and the subcarriers are randomly redistributed across the band. For the over-the-air testing, a single randomly-generated subcarrier placement is used for each value of $u$ and left unchanged for the duration of the testing.

\renewcommand{\arraystretch}{1.2}
\begin{table}[t]
\begin{center}
\caption{System parameters used for testing}
\label{tab:testParams}
\begin{tabular}{cccc}
\toprule[1pt]
Parameter 					& Symbol  		&  Value \\ \midrule
Bandwidth expansion		 	& $u$	 		& $1, 2, 4, 8$ \\
Number of subcarriers			& $K$			& $32$\\
Inserted pilot interval			& $\intvl$		& $4$\\
Symbols per second			& $f_\textup b$	& $1000$\\ 
Bits per symbol				& $B$			& $2$\\
Symbol alphabet				& $\bm \Theta$	& DFT matrix, $\F$\\
Subcarrier placement			& $F_\textup{nc}$ 	& Seg. random\\
\bottomrule[1pt]
\end{tabular}
\end{center}
\end{table}

The results presented in this section compare the average bit-error rate (BER) performance against the received SNR. The SNR is used instead of $E_\textup b/N_0$ to emphasize the low operating ranges of the developed systems. The received SNR is calculated as the ratio of received signal power divided by the noise power over the signal passband. That is, in a channel whose noise spectrum is flat across the transmission band, 
\begin{equation}\label{eqn:snr}
	\textup{SNR}=\frac{\int_{\textup{p.b.}} |Y_\textup{rx}(f)|^2~\dInt{f}}{\int_{\textup{p.b.}} |V(f)|^2~\dInt{f}},
\end{equation}
where $Y_\textup{rx}(f)$ denotes the received signal spectrum and $V(f)$ is the noise plus interference spectrum. The limits of the integrals, p.b., correspond to the passbands of the subcarriers, i.e., the frequency bands containing signal content.

\subsection{Channels with no interference}
\begin{figure}
	\centering
	\includegraphics[width=\figwid\columnwidth]{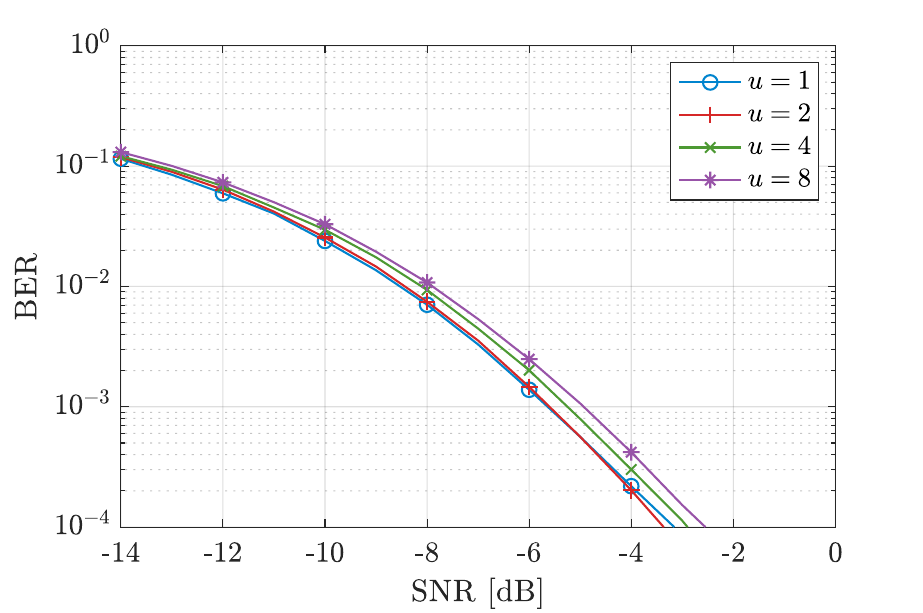}
	\caption{BER performance as $u$ varies in the MLD channel. The receiver is given perfect CSI.}
	\label{fig:mldFullCsi}
\end{figure}

To begin, the receiver is assumed to have perfect channel state information (CSI). In Fig.~\ref{fig:mldFullCsi}, we observe that the receiver performance degrades with increasing $u$. This loss is attributed to the spreading and subsequent truncation of the signal energy by the channel estimate when it is mapped to the contiguous domain through $\mb R$ in Fig.~\ref{fig:ncSymDet}. 

\begin{figure}
	\centering
	\includegraphics[width=\figwid\columnwidth]{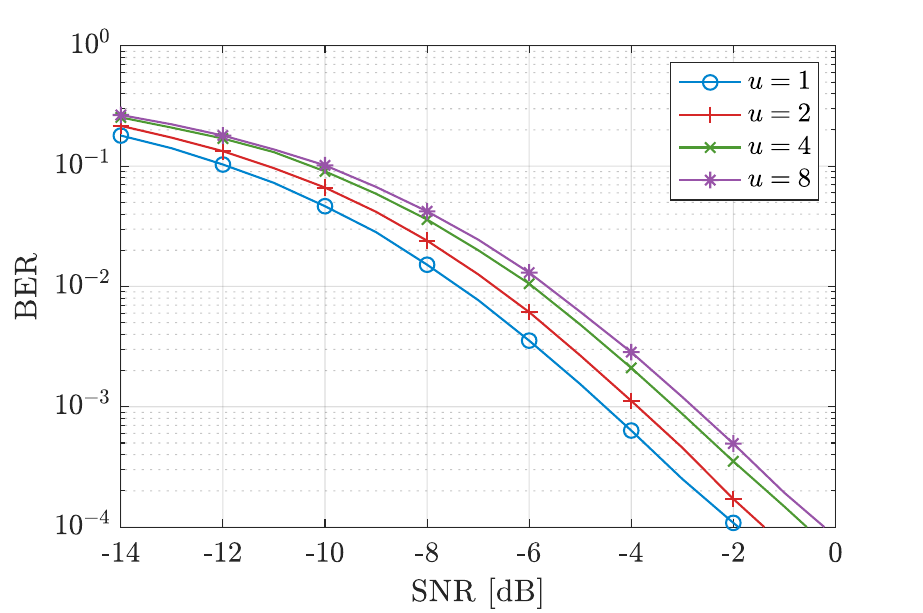}
	\caption{BER performance as $u$ varies in the MLD channel where each system must estimate the channel.}
	\label{fig:noCsi}
\end{figure}

Next, we consider the case where the receiver estimates the channel following the procedure detailed in Section~\ref{ssec:chan_acq}. The results of this test are given in Fig.~\ref{fig:noCsi}. Here, all cases of $u$ observe a performance loss due to channel estimation error, and the loss increases as the bandwidth expansion factor $u$ increases. 

\subsection{Channels with partial-band interference}
One immediate advantage of the NC-FMT-SS system is its ability to responsively allocate its signal energy subject to the presence of other users in the band. Whereas a contiguous system may encounter a certain level of unavoidable interference, a noncontiguous system can avoid interference by placing its subcarriers in the nulls between other users. We thus proceed by testing each of the systems in congested channel conditions under the same MLD channel used previously.

The channel congestion statistics used in this test are based on the data reported in \cite{warner_cad}: $25\%$ of a channel with bandwidth of $512$~kHz is polluted with interference. Specifically, $42$~users with a flat $3$~kHz bandwidth are randomly placed across the transmission band without overlap. Each user has a power-spectral density (PSD) $30$~dB greater than the average PSD of the signal over its passband. We assume that each system has perfect knowledge of the channel occupancy. The goal of each system is thus to allocate its signal energy to avoid as much interference as possible. For the contiguous system, the transmitter can only shift the center frequency of the waveform. For the noncontiguous systems, the transmitter can place its subcarriers to avoid interference provided they follow a segmented-random placement and the frequency extent $W_\tu e$ is less than $uW_\textup c$~Hz. 

\begin{figure}[t]
	\centering
	\includegraphics[width=\figwid\columnwidth]{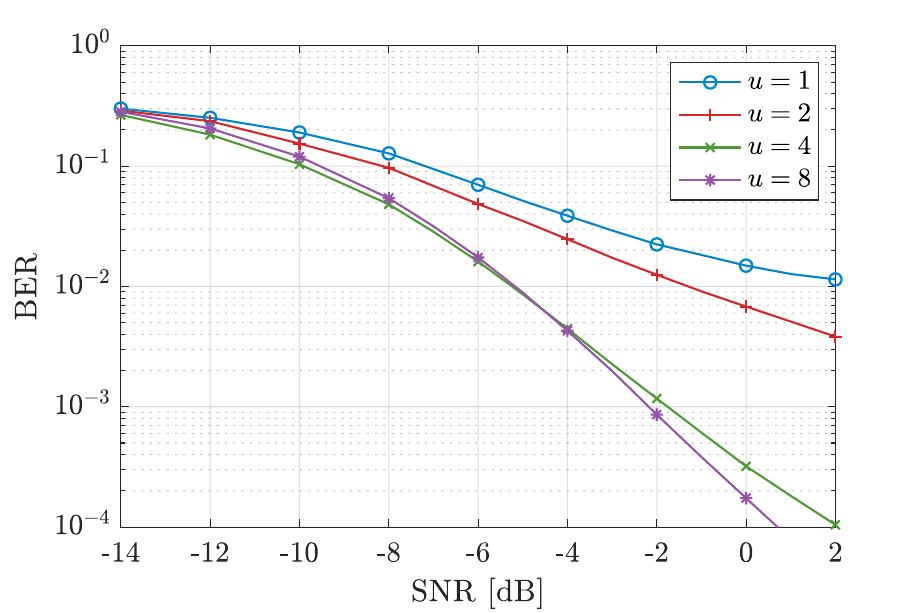}
	\caption{The performance of various NC-FMT-SS systems compared to C-FMT-SS in congested spectral conditions.}
	\label{fig:intf}
\end{figure}

The results of this test are presented in Fig.~\ref{fig:intf}. As the bandwidth expansion factor $u$ grows, the receiver has more freedom over where to place its subcarriers. Consequently, the wider-band received signals (e.g., $u=4$ and $u=8$) are received with less interference and thereby avoid the elevated BERs observed in the narrower-band systems. 

\subsection{Over-the-air results}
To demonstrate the viability of the developed system, we present performance results gathered in an over-the-air  (OTA) near-vertical-incidence skywave (NVIS) HF link. The transmitter was located in Idaho Falls, ID, and the receiver in Salt Lake City, UT. The testing was performed on October 3\textsuperscript{rd}, 2023, for $20$~minutes. 

A single packet for each value of $u$ was generated following the parameters detailed in Table~\ref{tab:testParams}. Each synthesized packet contained $512$~bits, and the timing phases were manually determined to avoid errors due to incorrect synchronization. 

\begin{figure}
	\centering
	\includegraphics[width=0.95\columnwidth]{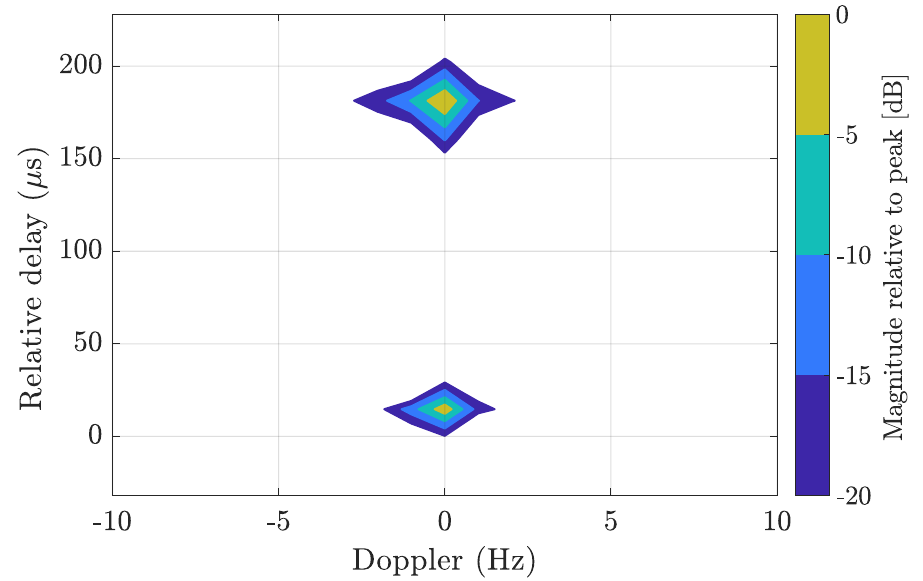}
	\caption{Estimated delay-Doppler domain channel response observed during the OTA testing. The delay axis is relative to the first identified multipath component.}
	\label{fig:otaChan}
\end{figure}

To obtain a high-SNR reference of the channel conditions during the testing, we periodically transmitted a $48$~kHz channel sounding sequence. Note that the receivers still performed channel estimation following Section~\ref{ssec:chanEst}; the sounding sequence is used only for reference. The resulting delay-Doppler domain representation of the channel response is given in Fig.~\ref{fig:otaChan}. This figure shows that the channel had two equal-power modes separated by roughly $170~\mu$s, and the $2\sigma$ Doppler spread was well below the $1$~Hz used to generate the simulation results. Because this channel condition is significantly milder than the MLD channel used to generate the simulated results, we conducted additional simulations based on these measured channel statistics to act as an expected performance curve for the OTA results. 

\begin{figure}
	\centering
	\includegraphics[width=\figwid\columnwidth]{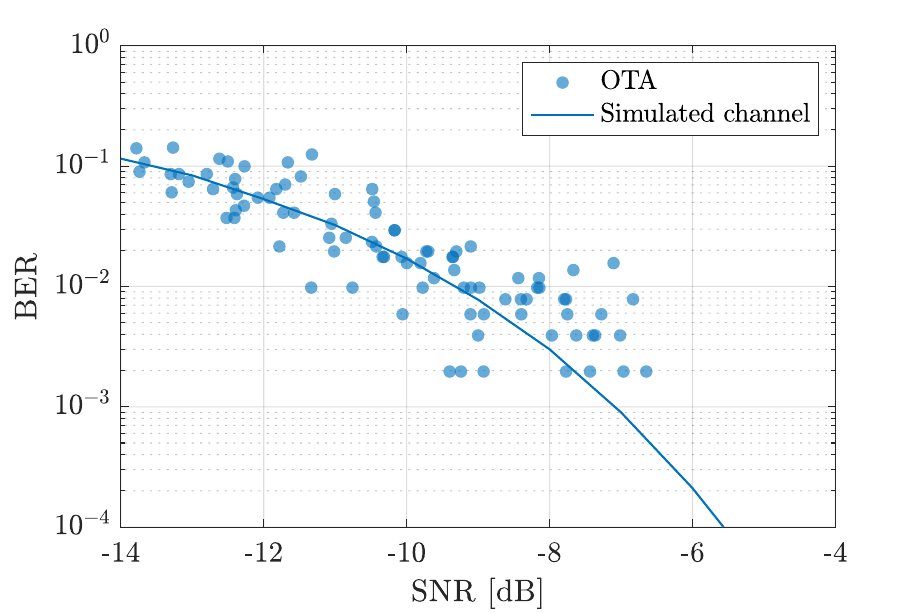}
	\caption{OTA results for the $u=1$ waveform. Each point represents the BER of an individually decoded packet. The simulated curve represents the performance obtained in a simulated channel based on the observed OTA channel conditions.}
	\label{fig:ota_u1}
\end{figure}

\begin{figure}
	\centering
	\includegraphics[width=\figwid\columnwidth]{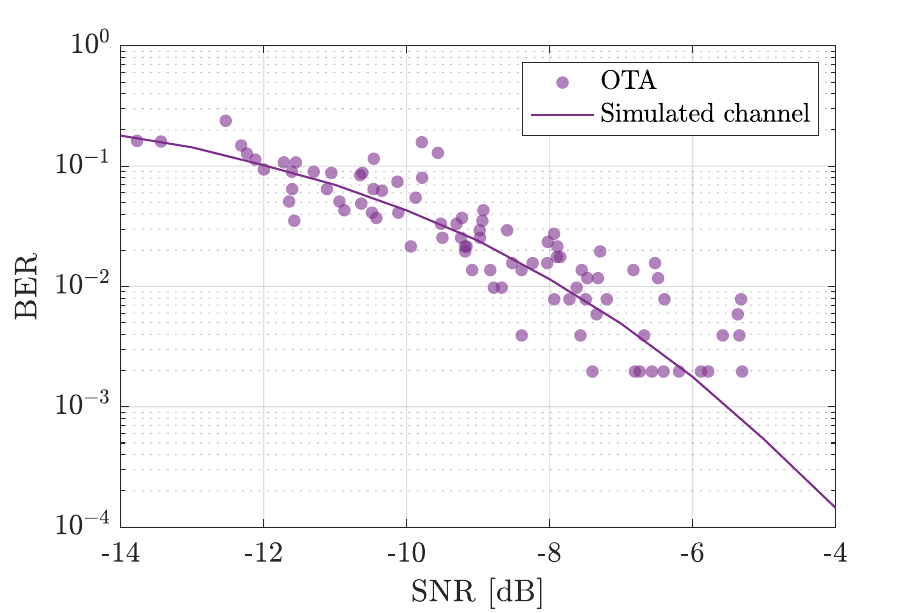}
	\caption{OTA results for the $u=8$ waveform. Each point represents the BER of an individually decoded packet. The simulated curve represents the performance obtained in a simulated channel based on the observed OTA channel conditions.}
	\label{fig:ota_u8}
\end{figure}

The performance of the $u=1$ and $u=8$ systems are given in Figs.~\ref{fig:ota_u1} and \ref{fig:ota_u8}, respectively. The results for the cases of $u=2$ and $u=4$ follow the same trends and are omitted for brevity. The scattered points indicate the BER of the individual packets. Note that the BER of the scattered points appears not to drop below $\approx 2\times 10^{-3}$ because each packet contains only $512$ bits (correspondingly, one bit error leads to a BER of $\approx 2\times 10^{-3}$). Many packets were recovered perfectly having a $\textup{BER}=0$ but are not shown on the logarithmic scale. 

\section{Conclusion}\label{sec:conclusion}
In this paper, we introduced and developed a noncontiguous filtered-multitone spread-spread (NC-FMT-SS) waveform. We examined several subcarrier placement methods, finding that a uniform placement yields the lowest PAPR, whereas random placements offer straightforward timing acquisition. In the receiver, we introduced a subcarrier-remapping step to reduce complexity, and we developed an estimator to recover the channel impulse response from its limited sampling. Lastly, we presented transmitter and receiver structures to facilitate practical implementations of these wideband signals.

The numerical results reveal that the developed NC-FMT-SS system has comparable performance (within roughly 1 to 2 dB) to a C-FMT-SS system in disturbed HF channels without interference. However, in congested spectral conditions, we found that the wider NC-FMT-SS systems ($u=4$ and $u=8$) significantly outperform the C-FMT-SS system, demonstrating performance gains of over $7$~dB at a BER of $10^{-2}$. Moreover, the over-the-air tests substantiate the simulated results, indicating the feasibility of the developed system in real-world channel conditions. Together, these results demonstrate that, although sufficient bandwidth may be unavailable to transmit wideband HF waveforms contiguously, robust performance can be achieved by adopting a noncontiguous subcarrier allocation.

%
\bibliographystyle{ieeetr}

\footnotesize
\bibliography{biblio}

\end{document}